\documentclass[prl,nobibnotes,twocolumn,showpacs]{revtex4}
\usepackage{graphics,color,array,dcolumn}
\usepackage{calc}
\usepackage[T1]{fontenc}
\usepackage[latin1]{inputenc}
\usepackage{amsmath}
\usepackage{amssymb}
\usepackage{xspace}
\allowdisplaybreaks[1]

\newcommand{\be}{\begin{equation}}
\newcommand{\ee}{\end{equation}}

\makeatother

\begin{document}
\title{Fixed points of Higher Derivative Gravity}
\author{Alessandro Codello}
\email{a.codello@gmail.com}
\affiliation{Dipartimento di Fisica Teorica, Universit\`a di Trieste, Viale Miramare, I-34014 Trieste, Italy}
\author{Roberto Percacci}
\email{percacci@sissa.it}
\affiliation{SISSA, via Beirut 4, I-34014 Trieste, Italy, and INFN, Sezione di Trieste, Italy}
\begin{abstract}
We recalculate the beta functions of higher derivative gravity in four dimensions
using the one--loop approximation to an Exact Renormalization Group Equation.
We reproduce the beta functions of the dimensionless couplings that were
known in the literature but we find new terms for the beta functions
of Newton's constant and of the cosmological constant.
As a result, the theory appears to be asymptotically safe at a 
non--Gaussian Fixed Point, rather than perturbatively renormalizable
and asymptotically free. 

\end{abstract} 
\maketitle

The earliest attempts at constructing a Quantum Field Theory (QFT) of gravity were
based on the application of perturbative methods to Einstein's theory.
It was soon understood that such methods would not succeed
due to the perturbative nonrenormalizability of Einstein's theory.
It was then natural to try with more general types of dynamics.
Lagrangians with four derivatives of the fields give propagators
that fall off with the fourth power of momentum, leading to improved
convergence of loop integrals.
It was indeed proven that a generalization of Einstein's theory containing 
terms quadratic in the curvature tensor is renormalizable in flat space 
perturbation theory [1].
It was also established in a series of papers [2-4] that the dimensionless
couplings of this theory (the inverse coefficients of the curvature squared terms)
are asymptotically free.
The beta functions of the dimensionful couplings -- Newton's constant $G$ and the
cosmological constant $\Lambda$ -- are gauge--dependent,
but the beta function of the dimensionless product $\Lambda G$ is not,
and this variable has also been claimed to be asymptotically free,
justifying the use of flat space perturbation theory.

A perturbatively renormalizable and asymptotically free QFT 
holds to arbitrarily high energy scales, so this could be regarded 
as a serious candidate for a fundamental theory of quantum gravity.
Unfortunately, it is not free of problems,
the most notorious one being the apparent lack of unitarity:
the ``bare'' action contains massive negative--norm states (ghosts) at tree level.
It was pointed out in [2,6] that these ghosts may not correspond to 
physical particles when quantum effects are taken into account,
but there exists to date no convincing proof that this happens.
Another, less well--known problem is that asymptotic freedom of $\Lambda G$ requires 
the choice of the unstable fixed point -5.467 for the parameter $\omega$
of eq.(1) below, see [5]. Pending progress on these issues,
higher derivative gravity does not seem to have gained wide acceptance
as a fundamental theory.
For a review of higher derivative gravity see [5]; for the state of the art see [7].

Nowadays, it is understood that Einstein's theory and its higher derivative
generalizations can be successfully treated as effective QFTs
with a cutoff presumably close to the Planck scale [8].
This is more than enough to cover all available experimental data,
but this fact has not stopped the search for an ``ultraviolet completion'' of the theory.
From the preceding remarks, it would seem that QFT
can only work if some nonperturbative mechanism is invoked,
the most promising one being as follows.
A QFT that admits a Fixed Point (FP) with a finite number of UV--attractive directions
can be predictive and hold to arbitrarily high energies. 
This behaviour was called ``asymptotic safety'' in [9].
A perturbatively renormalizable and asymptotically free theory is
a special case of asymptotically safe theory, where the FP is
the Gaussian FP (a free theory).
More general asymptotically safe theories will be based 
on nontrivial FPs.

The question arises whether a QFT of gravity could have this behavior.
The first positive evidence came long ago from studies in $2+\epsilon$ dimensions [9,10],
but technical issues then slowed down progress on this front for some time.
In the last ten years, using an Exact Renormalization Group Equation (ERGE),
the existence of a nontrivial FP has been established in four dimensions
for a truncation of the action containing the cosmological and 
Einstein--Hilbert terms [11,12], also in the presence of matter fields [13,14].
Independent evidence for a nontrivial FP also comes from Monte Carlo simulations [15,16].
However, so far only partial results are known for higher--derivative terms [17,18,14].

The behaviour of $\Lambda$ and $G$ in this approach is quite different
from the one predicted in the literature on higher derivative gravity.
In order to make a direct comparison,
we have recalculated the beta functions of higher derivative gravity,
starting from a one--loop approximation of the ERGE.
We find some important modifications in the beta functions
of Newton's constant and of the cosmological constant,
in such a way that the theory appears to be asymptotically safe at
a nontrivial FP, rather than at the Gaussian FP.
We report here the main results; details will be given elsewhere.

A general (Euclidean) theory containing terms quadratic in curvature has an action of the form
$$
\int d^4x\,\sqrt{g}\left[2 Z\Lambda-Z R
+\frac{1}{2\lambda}C^{2}-\frac{\omega}{3\lambda}R^{2}+\frac{\theta}{\lambda}E
\right]\, ,\eqno(1)
$$
where $Z=1/16\pi G$,
$C^{2}$ is the square of Weyl's tensor, $E$ is the integrand in 
Euler's topological invariant $\chi=\int dx\,\sqrt{g}E$.
We neglect the total derivative $\nabla^{2}R$.

For a quantum treatment, this action has to be supplemented by
the gauge--fixing term, which is chosen to be of the form
$$
S_{GF}=\int d^4x\sqrt{g}\,\chi_{\mu}Y^{\mu\nu}\chi_{\nu}
\eqno(2)
$$
where
$\chi_{\nu}=\nabla^{\mu}h_{\mu\nu}+\beta\nabla_{\nu}h$
(all covariant derivatives are with respect to the background metric) and
$Y^{\mu\nu}=\frac{1}{\alpha}\left[g^{\mu\nu}\nabla^{2}
+\gamma\nabla^{\mu}\nabla^{\nu}-\delta\nabla^{\nu}\nabla^{\mu}\right]$.
The ghost action contains the term
$$
S_c=\int d^4x\sqrt{g}\,\bar{c}_{\nu}(\Delta_{gh})_{\mu}^{\nu}c^{\mu}
\eqno(3)
$$
where
$(\Delta_{gh})_{\mu}^{\nu}=-\delta_{\mu}^{\nu}\square-(1+2\beta)\nabla_{\mu}\nabla^{\nu}+R_{\mu}^{\nu}$
as well as a term
$$
S_b=\frac{1}{2}\int d^4x\sqrt{g}\,b_{\mu}Y^{\mu\nu}b_{\nu}
\eqno(4)
$$
due to the fact that the gauge averaging operator $\mathbf{Y}$ depends nontrivially on the metric.
We follow earlier authors in choosing the gauge fixing parameters
$\alpha$, $\beta$, $\gamma$ and $\delta$
in such a way that the quadratic part of the action is:
$$
(\Gamma_k+S_{GF})^{^{(2)}}=\frac{1}{2}\int d^4x\sqrt{g}\ \delta g\mathbf{K}\mathbf{\Delta}^{(4)}\delta g
\eqno(5)
$$
where 
$\mathbf{\Delta}^{(4)}=\mathbf{1}\square^{2}+\mathbf{V}^{\rho\lambda}\nabla_{\rho}\nabla_{\lambda}+\mathbf{U}$.
For details of the operators $\mathbf{K}$, $\mathbf{V}$ and $\mathbf{U}$ 
we refer the reader to [7], whose notation we mostly follow.

The main tool in deriving nonperturbative information about the theory
is the gravitational ERGE [19]
\begin{widetext}
$$
\partial_{t}\Gamma_{k}=
\frac{1}{2}\mathrm{Tr}\left(
\frac{\delta^{2}(\Gamma_{k}+S_{GF})}{\delta g\delta g}
+\mathbf{R}_{k}^g\right)^{-1}
\!\!\partial_{t}\mathbf{R}_{k}^g
-\frac{1}{2}\mathrm{Tr}\left(
\frac{\delta^{2}S_b}{\delta b\delta b}
+\mathbf{R}_{k}^{b}\right)^{-1}
\partial_{t}\mathbf{R}_{k}^{b}
-\mathrm{Tr}\left(
\frac{\delta^{2}S_c}{\delta \bar c\delta c}
+\mathbf{R}_{k}^{c}\right)^{-1}
\partial_{t}\mathbf{R}_{k}^{c}
\, ,
\eqno(6)
$$
\end{widetext}
where $\Gamma_k$ is a coarse--grained effective action depending on a momentum scale $k$
and the kernels $\mathbf{R}_{k}$ act as infrared cutoffs.

In order to derive the beta functions of the couplings
$\tilde\Lambda=k^{-2}\Lambda$, $\tilde G=k^2 G$, $\lambda$, $\omega$ and $\theta$, 
we assume for $\Gamma_k$ the form (1) and insert it, together with the gauge--fixing and ghost terms (2,3,4),
into the ERGE.
Then, to calculate the r.h.s. of the ERGE we choose the cutoffs as follows:
$\mathbf{R}_k^g(\mathbf{\Delta}^{(4)})= \mathbf{K} R_k^{(4)}(\mathbf{\Delta}^{(4)})$,
$\mathbf{R}_k^c(\mathbf{\Delta}_{(gh)})=\mathbf{1}R_k^{(2)}(\mathbf{\Delta}_{(gh)})$,
$\mathbf{R}_k^b(\mathbf{Y})=\mathbf{1}R_k^{(2)}(\mathbf{Y})$,
where 
$R_k^{(n)}(z)$ is a suitable profile function chosen to
suppress the propagation of field modes with momenta below $k$.
We will use the so--called optimized cutoff [20] $R_k^{(n)}(z)=(a k^n-z)\theta(a k^n-z)$,
with $a=1$ unless otherwise stated.

We restrict ourselves to the one--loop approximation, 
which in the context of the ERGE consists of taking into account only the explicit dependence
of $R_k(z)$ on $k$, neglecting the implicit dependence due to the presence
of running couplings in the cutoff function.
(In the case of the Einstein--Hilbert action, where the r.h.s. of the ERGE can
be computed exactly, it is known that this approximation does not 
change the general behaviour.) The traces are evaluated with heat kernel methods,
keeping all terms up to $B_4$, and using the results of [21].
This procedure provides a logically and computationally independent derivation
of the beta functions.

The beta functions of the dimensionless couplings appearing in (1) turn out to be:
\begin{align*}
\beta_{\lambda} & =  -\frac{1}{(4\pi)^{2}}\frac{133}{10}\lambda^{2}\ ,\tag{7a}\\
\beta_{\omega} & =  -\frac{1}{(4\pi)^{2}}\frac{25 + 1098\,\omega+ 200\,\omega^2}{60}\lambda\ ,\tag{7b}\\
\beta_{\theta} & =  \frac{1}{(4\pi)^{2}}\frac{7(56-171\,\theta)}{90}\lambda\ .\tag{7c}
\end{align*}
They agree with those calculated in dimensional regularization [4,5,7]. 
The coupling $\lambda$ has the usual logarithmic approach
to asymptotic freedom, while the other two couplings have the FP values
$\omega_*\approx(-5.467,-0.0228)$ and
$\theta_*\approx0.327$.
Of the two roots for $\omega$, the first turns out to be UV--repulsive, 
so the second has to be chosen [4,5,7].

The beta functions of 
$\tilde\Lambda$ and $\tilde G$ are:
\begin{widetext}
\begin{align*}
\beta_{\tilde \Lambda} & =  
-2\tilde\Lambda
+\frac{1}{(4\pi)^{2}}\left[
\frac{1+20\omega^2}{256\pi\tilde G\omega^2}\lambda^2
+\frac{1+86\omega+40\omega^2}{12\omega}\lambda\tilde\Lambda\right]
-\frac{1+10\omega^2}{64\pi^2\omega}\lambda
+\frac{2\tilde G}{\pi}
-q(\omega)\tilde G \tilde\Lambda \tag{8a}\\
\beta_{\tilde G} & =  2\tilde G
-\frac{1}{(4\pi)^{2}}\frac{3+26\omega-40\omega^2}{12\omega}\lambda\tilde G
-q(\omega) \tilde G^2
\,.\tag{8b}
\end{align*}
\end{widetext}
where $q(\omega)=(83+70\omega+8\omega^2)/18\pi$.
The first two terms in each beta function exactly 
reproduce the results of [4,5,7], the remaining ones are new.
The origin of the new terms can be easily understood.
The beta functions were originally derived as coefficients of $1/\epsilon$ 
poles in dimensional regularization, which correspond to logarithmic divergences 
in the effective action.
In our heat kernel derivation these terms are given by the $B_4$ coefficient.
The new terms that we find come from the $B_2$ and $B_0$ coefficients, 
which in a conventional calculation of the effective action would correspond 
to quadratic and quartic divergences. 
Dimensional regularization is ill--suited to compute these terms.
It is important to stress that our ``Wilsonian'' calculation of the beta functions
does not require any UV regularization. The only ambiguity is in the
choice of the cutoff functions, but no reasonable choice
could remove the $B_2$ and $B_0$ terms.

To picture the flow of $\tilde\Lambda$ and $\tilde G$,
we set the remaining variables to their FP values $\omega=\omega_*$, $\theta=\theta_*$, 
and $\lambda=\lambda_*=0$. Then, defining $q_*=q(\omega_*)\approx 1.440$
the flow equations (8) can be solved analytically:
\begin{align*}
\tilde \Lambda(t)&=
\frac{(2\pi\tilde\Lambda_0-\tilde G_0(1-e^{4t}))e^{-2t}}
{\pi(2-q_*\tilde G_0(1-e^{2t}))}\tag{9a}\ ,\\
\tilde G(t)&=
\frac{2 \tilde G_0 e^{2t}}
{2-q_*\tilde G_0(1-e^{2t})}\ .\tag{9b}
\end{align*}

The resulting flow in the $(\tilde\Lambda,\tilde G)$--plane is
shown in Fig.1.
\begin{figure}[t]
    \center{\resizebox{1\columnwidth}{!}{\includegraphics{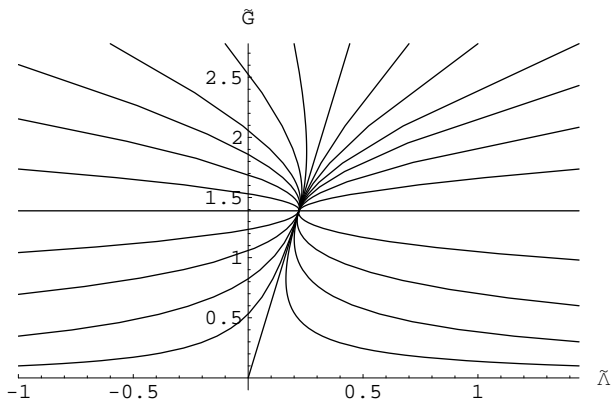}}
    \caption{\label{Fig:1} The flow in the $(\tilde\Lambda,\tilde G)$--plane}}
\end{figure}
It has two FPs: the Gaussian FP at $\tilde{\Lambda}=\tilde{G}=0$ and another one at
$$
\tilde{\Lambda}_{*}=\frac{1}{\pi q_*}\approx 0.221\ ,\ \ \ \ \ 
\tilde{G}_{*}=\frac{2}{q_*}\approx 1.389\ .
\eqno(10)
$$
The attractivity properties of these FPs are determined by the stability matrix 
$$
M_{ij}=\frac{\partial\tilde{\beta}_{i}}{\partial\tilde{g}_{j}}=
\left( \begin{array}{cc}
-2-q_*\tilde G & \frac{2}{\pi}-q_*\tilde\Lambda  \\
0 & 2-2q_*\tilde G  \end{array}\ . \right)
$$
At the Gaussian FP the eigenvalues of $M$ are $(-2,2)$;
the attractive eigenvector points along the $\tilde\Lambda$ axis and 
the repulsive eigenvector has components $(1,2\pi)$.
As expected on general grounds [9], the eigenvalues are
the opposite of the canonical dimensions of $\Lambda$ and $G$.
At the non--Gaussian FP the eigenvalues of $M$ are $(-4,-2)$
with the same eigenvectors as before.
The FP given by (10) is UV--attractive in all five couplings.
Note the ``critical'' trajectory
joining the Gaussian to the non-Gaussian FP, 
which is tangent to the repulsive eigenvector in the origin and is actually given by
$\tilde G(t)=2\pi\tilde\Lambda(t)$ for all $-\infty<t<\infty$.

From this calculation one can derive some physical predictions.
The first is the UV--limit of the cosmological
constant in Planck units $\Lambda G=\tilde\Lambda\tilde G$,
which is well known to be gauge--independent
and is also independent of the cutoff parameter $a$.
In contrast to [4,5,7], we find that $\Lambda G$ tends 
to the finite value $2/(\pi q_*^2)\approx 0.307$.
Of course this is an asymptotic UV value 
and to compare it with cosmological observations
one would have to run the RG down to extremely low values of $k$.

Another prediction is the asymptotic value $-2\omega_*/3\approx 0.0152$ 
for the ratio between the coefficients of $R^2$ and $C^2$.
It is interesting to observe that the flow induced by a large
number $N$ of minimally coupled matter fields gives for this ratio the value
$5n_S/(3n_S+18 n_D+36n_M)$, where $n_S$, $n_D$ and $n_M$
are the numbers of scalar, Dirac and gauge fields [14].
This number is also quite small in realistic unified theories.
Thus both with and without matter it seems that, in the UV limit, 
fluctuations of the conformal tensor
will be more suppressed than fluctuations of the Ricci tensor.

The flow that we find here is almost identical to the flow
obtained in the large $N$ limit [14],
where the coefficients $\omega_*$, $\theta_*$ and $q_*$
are determined by $n_S$, $n_D$ and $n_M$.
A remarkable feature of the large $N$ limit, in conjunction with the use of
optimized cutoffs, is that all higher powers of curvature are absent at the FP.
This raises the hope that asymptotically safe gravity may be
describable by a finite number of terms in the action
(generically, one would expect to have infinitely many terms, with relations 
between the coefficients such that only a finite number of parameters is left arbitrary).

Our flow is also similar to the one obtained in the Einstein--Hilbert truncation [12],
where, however, the critical exponents at the non-Gaussian FP are complex, 
resulting in a spiralling approach to the FP.
This similarity may be somewhat surprising, because in the Einstein--Hilbert truncation
the higher derivative terms are absent while here they dominate the dynamics.
To some extent it can be understood by the following argument.
In gravity at low energies the couplings do not run, and therefore the relative
importance of the terms in the action can be determined simply by
counting the number of derivatives of the metric.
For example, at low momenta $p\ll \sqrt{Z}$ (recall that $Z$ is the square of the Planck mass), 
the terms in the action (1) with four derivatives are
suppressed relative to the term with two derivatives by a factor $p^2/Z$.
This is not the case in the FP regime:
if we consider phenomena occurring at an energy scale $p$,
then also the couplings should be evaluated at $k\approx p$.
If there is a nontrivial FP, $Z$ runs exactly as $p^2$
and therefore both terms are of order $p^4$.
This is just a restatement of the fact that in the Einstein--Hilbert
truncation the graviton has an anomalous dimension equal to two,
making its propagator behave like $p^{-4}$ at high energy.

Partial results for the four--derivative couplings,
but going beyond one loop, have been derived using the ERGE in [17].
Using a spherical background,
where $\int d^4x\sqrt{g}C^2=0$, $\int d^4x\sqrt{g}R^2=384\pi^2$ and $\chi=2$,
the beta function of the combination
$-\frac{\omega}{3\lambda}+\frac{1}{192\pi^2}\frac{\theta}{\lambda}$
can be derived.
In the absence of further input it is impossible to
disentangle the beta functions of the individual couplings.
Nevertheless, this provides valuable information.
In particular, since a finite FP--value is found for a combination
of couplings, this calculation suggests that the asymptotic freedom
of $\lambda$, $\lambda/\omega$ and $\lambda/\theta$
that we find here may be only a consequence of the approximations that we made,
and that in a more accurate calculation some or all of these coefficients 
will reach finite values instead of running logarithmically.
One also expects, as in [17], that the degeneracy 
of the stability matrix is lifted
and that all couplings are either relevant or irrelevant.

To summarize, we have found that within our approximations
higher derivative gravity has a fixed point with the
following properties:
$\tilde\Lambda$ and $\tilde G$ are nonzero and UV--relevant,
while the couplings of the terms quadratic in curvature are 
asymptotically free and marginal.
Experience with the Einstein--Hilbert truncation suggests that the FP
will persist in a more precise treatment, up to a finite shift of
the FP--values of the couplings, and of the critical exponents.
The Gaussian FP is unstable: even an infinitesimal value for $\tilde G$
will generate a nonvanishing $\tilde\Lambda$ and push the system
towards the other FP.

Among other things, these results solve the second of the problems 
mentioned in the introduction.
Concerning the issue of unitarity, we can say,
from our Wilsonian point of view, that the
presence of ghost poles at the Planck scale has to be assessed
by considering the action $\Gamma_k$ for $k\approx m_{\rm Planck}$,
which is probably quite different from the FP action.
Thus, tree level analyses of the FP action are of little significance,
as already pointed out in [2,3,6].
This is generally accepted in the case of QCD:
a tree level analysis of the QCD FP action would predict the existence 
of states that are not observed in the physical spectrum,
but this is no longer considered a serious argument against this theory.
In view of this, and of the results reported here, 
we think that higher derivative gravity deserves renewed attention.

\bigskip
\centerline{\bf Acknowledgements}
RP would like to thank M. Reuter for useful conversations. 
We would also like to thank G. de Berredo--Peixoto
and I. Shapiro for correspondence on their work.
\goodbreak
\medskip

\centerline{\bf References}

\noindent [1] K.S. Stelle,
Phys. Rev. {\bf D16}, 953 (1977).

\noindent [2] J. Julve, M. Tonin, Nuovo Cim. {\bf 46B}, 137 (1978).

\noindent [3] E.S. Fradkin, A.A. Tseytlin, 
Phys. Lett. {\bf 104 B}, 377 (1981).
Nucl. Phys. {\bf B 201}, 469 (1982).

\noindent [4] I.G. Avramidi, A.O. Barvinski, 
Phys. Lett. {\bf 159 B}, 269 (1985).

\noindent [5] I.L. Buchbinder, S.D. Odintsov and I.L. Shapiro, 
``Effective action in quantum gravity'',
IOPP Publishing, Bristol (1992).

\noindent [6] A. Salam, J. Strathdee, Phys.Rev.{\bf D18} 4480 (1978).

\noindent [7] G. de Berredo--Peixoto and I. Shapiro, Phys.Rev. {\bf D71} 064005 (2005);
[arXiv:hep-th/0412249].

\noindent [8] Cliff P. Burgess,
Living Rev. in Rel. 7,  (2004),  5; 
J.F. Donoghue, [arXiv:gr-qc/9512024].

\noindent [9] S. Weinberg, 
In {\it General Relativity: An Einstein centenary survey}, 
ed. S.~W. Hawking and W. Israel, pp.790--831; 
Cambridge University Press (1979).

\noindent [10] S.M. Christensen and M.J. Duff, {\it  Phys. Lett.} {\bf B 79}, 213 (1978).
R. Gastmans, R. Kallosh and C. Truffin, {\it  Nucl.\ Phys.} {\bf B 133}, 417 (1978).

\noindent [11] W. Souma, Prog. Theor. Phys. {\bf 102}, 181 (1999); [arXiv:hep-th/9907027].

\noindent [12] O. Lauscher and M. Reuter, Phys. Rev. {\bf D65}, 025013 (2002);
[arXiv:hep-th/0108040];
Class. Quant. Grav. {\bf 19}, 483 (2002);
[arXiv:hep-th/0110021]; 
Int. J. Mod. Phys. {\bf A 17}, 993 (2002);
[arXiv:hep-th/0112089];
M. Reuter and F. Saueressig, Phys. Rev. {\bf D65}, 065016 (2002);
[arXiv:hep-th/0110054].

\noindent [13] R. Percacci and D. Perini, Phys. Rev. {\bf D67}, 081503(R) (2003),
[arXiv:hep-th/0207033]; Phys. Rev. {\bf D68}, 044018 (2003),
[arXiv:hep-th/0304222].

\noindent [14] R. Percacci , Phys. Rev. {\bf D73}, 041501(R) (2006),
[arXiv:hep-th/0511177].

\noindent [15] J. Ambj\o rn, J. Jurkiewicz, R. Loll, 
Phys. Rev. Lett. {\bf 95} 171301 (2005),
[arxiv:hep-th/0505113]; 
Phys. Rev. {\bf D72} 064014 (2005),
[arxiv:hep-th/0505154].

\noindent [16] H. Hamber and R. Williams (2004), Phys. Rev. {\bf D 70}, 124007 (2004);
[arXiv:hep-th/0407039].

\noindent [17] O. Lauscher and M. Reuter, Phys. Rev. {\bf D 66}, 025026 (2002).

\noindent [18] A.A. Bytsenko, L.N. Granda, S.D. Odintsov,
JETP Lett.65:600-604 (1997) [arXiv:hep-th/9705008];
L.N. Granda, S.D. Odintsov, Grav.Cosmol.4:85-95 (1998) [arXiv:gr-qc/9801026].

\noindent [19] M. Reuter, Phys. Rev. {\bf D57}, 971 (1998).

\noindent [20] D.F. Litim, Phys.Rev. {\bf D 64} 105007 (2001),
[arXiv:hep-th/0103195]; 
Phys.Rev.Lett. {\bf 92} 201301 (2004), [arXiv:hep-th/0312114].

\noindent [21] V.P. Gusynin, Nucl. Phys. {\bf B 333}, 296-316 (1990);
V.P. Gusynin and V.V. Kornyak, Nucl. Instr. and Meth. in Phys. Res. {\bf A 389}, 365-369 (1997).
V.P. Gusynin and V.V. Kornyak, [arXiv:math.NA/9909145].

\end{document}